\documentclass[iop]{emulateapj}
\newcommand\sqd{{deg$^{2}$}}
\newcommand\kms{km~s$^{-1}$}
\newcommand\msun{$M_\odot$}

\newcommand{\etal}{et al.}

\def\be{\begin{equation}}
\def\ee{\end{equation}}

\usepackage{amsmath}

\shorttitle{ALFALFA Local Volume Galaxy}
\shortauthors{R. Giovanelli et al.}

\begin{document}
\title{ALFALFA Discovery of the Nearby Gas-Rich Dwarf Galaxy Leo~P. \\  I. HI Observations}
\author {Riccardo Giovanelli\altaffilmark{1}, Martha P. Haynes\altaffilmark{1}, 
Elizabeth A.K. Adams\altaffilmark{1}, John M. Cannon\altaffilmark{2}, 
Katherine L. Rhode\altaffilmark{3}, John J. Salzer\altaffilmark{3}, Evan D. Skillman\altaffilmark{4}, 
Elijah Z. Bernstein-Cooper\altaffilmark{2} and  Kristen  B. W. McQuinn\altaffilmark{4} }

\altaffiltext{1}{Center for Radiophysics and Space Research, Space Sciences Building,
Cornell University, Ithaca, NY 14853. {\it e--mail:} riccardo@astro.cornell.edu,
haynes@astro.cornell.edu, betsey@astro.cornell.edu}

\altaffiltext{2}{Department of Physics and Astronomy,
Macalester College, Saint Paul, MN 55105. {\it e--mail:} jcannon@macalester.edu, ebernste@macalester.edu}

\altaffiltext{3}{Department of Astronomy, Indiana University, Bloomington, IN 47405. 
{\it e--mail:} rhode@astro.indiana.edu, slaz@astro.indiana.edu}

\altaffiltext{4}{Minnesota Institute for Astrophysics,
University of Minnesota, Minneapolis, MN 55455. {\it e--mail:} skillman@astro.umn.edu,
kmcquinn@astro.umn.edu}

\begin{abstract}
The discovery of a previously unknown 21cm HI line source identified as 
an ultra--compact high velocity cloud in the ALFALFA survey is reported. 
The HI detection is barely resolved by the Arecibo 305m telescope $\sim$4\arcmin\ 
beam and has a narrow HI linewidth (HPFW of 24 \kms). Further HI observations 
at Arecibo and with the VLA corroborate the ALFALFA HI detection, provide 
an estimate of the HI radius,
$\sim 1$\arcmin~ at the $5\times 10^{19}$ cm$^{-2}$ isophote, and show the 
cloud to exhibit rotation with an amplitude of $\simeq$ 9.0$\pm$1.5 \kms.
In other papers, \citet{Rhode2013} show the HI source to have a resolved 
stellar counterpart and ongoing star forming activity, while \citet{Skillman} 
reveal it as having extremely low metallicity: $12+\log (O/H)=7.16\pm 0.04$. 
The HI mass to stellar mass ratio of the object is found to be 2.6. We use 
the Tully--Fisher template relation in its baryonic form \citep{McGaugh} to 
obtain a distance estimate $D_{Mpc}=1.3^{+0.9}_{-0.5}$. Additional constraints
on the distance are also provided by the optical data of \citet{Rhode2013} 
and McQuinn et al. (private communication), both indicating a distance in the 
range of 1.5 to 2.0 Mpc. The three estimates are compatible within their errors. 
The object appears to be located beyond the dynamical boundaries of, but still 
in close proximity to the Local Group. Its pristine properties are consistent 
with the sedate environment of its location. At a nominal distance of 1.75 Mpc, 
it would have an HI mass of $\simeq 1.0\times 10^6$ \msun, a stellar mass of 
$\simeq 3.6\times 10^5$ \msun, and a dynamical mass within the HI radius of 
$\simeq 1.5\times 10^7$ \msun. This discovery 
supports the idea that optically faint --- or altogether dark --- low mass halos
may be detectable through their non-stellar baryons.
\end{abstract}

\keywords{galaxies: spiral --- galaxies: distances and redshifts ---
galaxies: halos --- galaxies: luminosity function, mass function ---
galaxies: photometry --- radio lines: galaxies}

\section {Introduction}\label{intro}

The faint end of the luminosity function of galaxies is generally 
modeled by a power law which is significantly shallower than that 
prescribed by the $\Lambda$CDM scenario for the low end of 
the mass function of dark matter halos. When this mismatch between
observations and theory is found in the vicinity 
of massive galaxies such as the Milky Way (MW),  it is 
referred to as the ``missing satellite problem'' \citep{Klypin1999}; 
when encountered in regions of very low galaxy density, it is 
referred to as the ``void phenomenon'' \citep{Peebles2001}. Here, we 
shall refer to the issue generically as the ``dwarf galaxy problem'': 
the observed underabundance of low-mass galaxies relative to theoretical
expectations. The conflict between observations and theory is also 
encountered in the behavior of the HI mass function \citep{Martin2010}, 
the rotational velocity width function \citep{Papastergis2011}, and the 
baryonic mass function of galaxies \citep{Papastergis2012}. The numerous 
ultra-faint dwarf galaxies found in the vicinity of the MW and M31 
discovered over the last decade in wide-field optical surveys have 
alleviated somewhat --- but by no means resolved --- the mismatch between 
theory and observations \citep{Koposov2008}. Moreover, the ultra-faint 
dwarf satellites of the MW are located at small galactocentric 
distances, well inside the virial radius of the MW, and thus their 
structural and evolutionary properties have been drastically affected 
by tidal and hydrodynamical processes \citep{Sand2012}, making it difficult 
to assess their primitive characteristics.

The possibility that HI observations might help resolve the dwarf galaxy 
problem was proposed in two influential papers more than a decade ago
\cite[]{Blitz1999, Braun1999}, in which the HI high velocity clouds (HVCs) 
were invoked as baryonic but starless counterparts to low-mass dark 
matter halos in the Local Group. In particular, the more compact 
of those HVCs (sizes $\lesssim 2^\circ$)
were proposed as representative of the relatively primordial, yet 
undisturbed population of LG low-mass halos. This interpretation was 
challenged by \citet[hereafter SMW02]{Sternberg2002}, on the grounds that, 
if placed at typical LG distances ($\sim 1$ Mpc): (i) the sizes of even the more 
compact HVCs (CHVCs) known at the time were too large 
(sizes $>$ 10 kpc) to match the mass--concentration 
relation of $\Lambda$CDM halos; and (ii) the CHVCs would be
too massive ($M_{HI}>10^7$ \msun)
to explain the lack of detection of similar populations in nearby groups of 
galaxies \citep{Pisano2007}. The existence of gas-bearing but starless (or 
nearly so) minihalos was recently revived by \citet[][hereafter G+10]{Giov2010} 
who identified a category of even more compact HVCs (sizes $\lesssim 10\arcmin$, 
thus the ultra--compact HVCs: UCHVCs) in the ALFALFA extragalactic HI
survey \citep{Giov2005}. The thermal models of SMW02 are given for a broad 
range of intergalactic medium (IGM) pressures. While we don't know the specific
environmental conditions at the location of any single cloud, plausible models 
can be fit to the UCHVC data. 
Placed to distances of a few Mpc, their HI masses would be smaller than 
detection limits of extant surveys of nearby groups. The one critical item 
preventing the validation of the idea that the UCHVCs can be counterparts 
of extragalactic minihalos is that their distances are not known. There are two 
main routes to advancing such validation; one would be the detection 
of stellar populations associated with the UCHVCs; the other would be the detection 
of a population of objects with comparable HI properties in a nearby 
group of galaxies, clearly associated by virtue of similarity in the 
distribution of radial velocities. Here we report on a finding that  
allows for progress along the former route. The object involved is a dwarf galaxy 
located in the immediate vicinity of the LG.

\section{Discovery of HI102145.0+180501}\label{alfalfa}

Making use of the seven-horn Arecibo L-band Feed Array (ALFA), the
Arecibo Legacy Fast ALFA Survey ALFALFA \citep{Giov2005} is a blind survey 
in the HI 21cm line covering 7000 \sqd~  of high Galactic latitude sky 
in two regions: a ``Spring'' northern Galactic circumpolar 
region between  $7.5h$ and $16.5h$ in R.A. and a ``Fall'' southern Galactic 
region between $22h$ and $03h$ in R.A. Both footprints extend from $0^\circ$ 
to $36^\circ$ in Declination. The spectral coverage extends from 
$-2000$ to $+18000$ \kms~ with a channel separation of $\sim$5.5 \kms. 
Initiated in 2005 and completed in 2012,
survey observations were carried out with the 
305~m telescope at the Arecibo Observatory
\footnote{The Arecibo Observatory is operated by 
SRI International under a cooperative agreement with the 
National Science Foundation (AST-1100968), and in alliance with Ana G. 
M\'endez-Universidad 
Metropolitana and the Universities Space Research Association.}
with an angular resolution of $\sim 4$\arcmin~ and an
integration time of $\sim$40 sec per beam, yielding an r.m.s. 
flux sensitivity of $\sim 2.5$ mJy  for a spectral feature smoothed
to 11 \kms. A catalog containing sources over nearly 3000 \sqd~ (40\% of the 
total ALFALFA sky) is in the public domain \citep{Haynes2011}.

As discussed in previous works (Haynes et al. 2011 and references therein), the ALFALFA
survey observations consist of a series of drift scans which are combined
to construct 3-D data cubes. Data processing is carried out sequentially
on stripes $2^\circ.4$ wide in Declination. Signal identification is performed by
applying a matched filtering algorithm in the Fourier domain \citep{Saintonge2007}.
The process of catalog construction also includes the inspection of 
optical databases and the assignment of the most probable optical counterpart
to each HI signal.

During the course of construction of the ALFALFA catalog
along the Spring stripe centered at Dec.= +17$^\circ$,
an HI source with heliocentric velocity of 264 \kms~ was detected at 
$102145.0+180501$~(epoch 2000.0).
The source was designated as an UCHVC of the type reported by G+10.
The HI emission appeared to be very weakly, if at all, resolved by 
the 4\arcmin~ ALFA beam. Although no cataloged optical galaxy is 
recorded near the HI position, a very faint blue object appears in both the 
SDSS and DSS2 blue images, $\sim$ 20\arcsec~ NE of the ALFALFA centroid, but 
within the positional error box. The SDSS image exhibits an irregular, lumpy light
distribution that could be described as being marginally resolved into
stars. In fact, what we now identify 
as a single galaxy is seen as a combination of several photometric objects 
in the SDSS database, some of which are classified as stars and some as 
galaxies. It is identified as a compact group of galaxies (SDSSCGB11269) by
\citet{McConnachie2009} who used an automated search algorithm to identify
such objects in the SDSS DR6 catalog. As those authors note, spurious 
identification of compact groups can be due to errors in the photometric catalog.
In fact, this group candidate is included in their less reliable ``catalog B''.
Aided by the good position match with that of the HI source, however, the 
visual inspection of both the SDSS and the DSS2 blue images suggested an 
alternative interpretation: that of a very faint, nearby, star forming dwarf 
galaxy. Follow-up optical observations 
reported by \citet{Rhode2013}: (i) confirmed that the HI source
has a stellar counterpart, 
(ii) provided an accurate optical flux, partly resolved into individual stars,
(iii) detected H$\alpha$ emission indicative of current star formation activity 
and (iv) produced an independent estimate of the galaxy's distance. Later optical 
spectroscopy of the \ion{H}{2} region 
revealed the galaxy's extremely low metallicity
\citep{Skillman}. We shall hereafter refer to it as Leo P because of its 
apparently pristine nature. It is also identified as AGC208583 in the
catalog maintained by MPH and RG and reported in a number of on-line 
extragalactic data bases.

\section{Properties of the Galaxy Leo P}\label{LeoP}

The follow-up observations mentioned above include BVR broad-band imaging obtained 
in excellent seeing conditions (0.6\arcsec--0.8\arcsec) with the  WIYN 3.5-m
telescope and more recently with the Large Binocular telescope (LBT); 
H$\alpha$ imaging with the 2.1m KPNO telescope; optical spectroscopy with the 
4m KPNO telescope and the Large Binocular Telescope (LBT); and 
HI synthesis imaging with the  Karl G. Jansky Very Large Array
({\it VLA}\footnote{The National Radio Astronomy Observatory is a
  facility of the National Science Foundation operated under
  cooperative agreement by Associated Universities, Inc.}).
Ongoing fitting of  thermal structure models will yield insights into the phases
of Leo P's interstellar medium (Faerman \etal, in preparation). Throughout this
paper, properties of Leo P will be tabulated through their explicit dependence
on the object's distance $D_{Mpc}$ in Mpc. As we discuss in Sections 3.1 and 3.3, that
distance is likely to be between 1.5 and 2 Mpc.

\subsection{Summary of Optical Properties}

Using the WIYN telescope broad-band imaging data, \citet{Rhode2013} clearly detected 
the stellar population associated with Leo P, showed that it extends to a radius of 
$\simeq 225 D_{Mpc}$ pc and has a stellar mass of $1.2\times 10^5 ~D_{Mpc}^2$ \msun. 
They also obtained constraints to the distance $D_{Mpc}$ of Leo P. After extraction 
of a sample of $\simeq 10^2$ resolved stars, a color-magnitude diagram allowed an
identification of the red giant stellar branch (RGB), to which isochrones could
be fitted. However, the sparseness of the stellar population, coupled with the limited
depth of their photometric data, makes the determination of the location of the 
branch tip (TRGB) difficult. Recently, deeper optical broad band images of the Leo P 
field have been obtained with the LBT by K. McQuinn et al. (private communication). 
These data allow a clear identification of the RGB but, again, due to the 
sparseness of the resolved stellar population, the location of the TRGB cannot yet 
be sharply defined. Both groups do however agree that the analysis of their 
respective data sets indicate that the distance to Leo P lies 
somewhere between 1.5 and 2.0 Mpc. These estimates are complemented by, and in 
agreement with that of $D_{Mpc}=1.3^{+0.9}_{-0.5}$ obtained using 
the baryonic Tully-Fisher relation, presented in Section 3.3. Optical 
spectroscopy data of \citet{Skillman} obtained with the LBT shows Leo P
to have the extremely low metallicity of 12 + log(O/H)=7.16$\pm0.04$, and to 
exhibit essentially primordial Helium abundance.

\begin{figure}
\figurenum{1}
\begin{center}
\includegraphics[width=\linewidth]{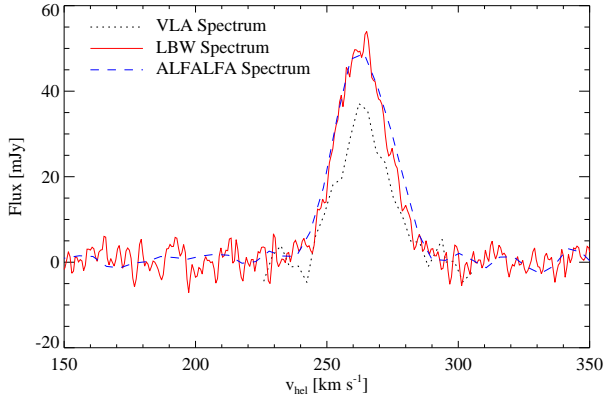}
\caption{HI line spectra of Leo P from separate observations:
the original ALFALFA spectrum (dashed blue line), the higher resolution
spectrum from the targeted LBW observation (solid red line), and the spatially
integrated spectrum extracted from the masked VLA cube used to create the moment map 
in  Fig. \ref{fig:f2} (dotted black line). The LBW observation matches the ALFALFA 
spectrum very closely and, despite its substantially higher spectral resolution, 
it does not reveal any additional spectral detail. The VLA integrated spectrum,
sampled at 3.3 \kms, shows missing flux relative to the single-dish observations.}
\end{center}
\label{fig:f1}
\end{figure}

\subsection{HI Properties}

Figure \ref{fig:f1} displays the HI spectral profile of Leo P as extracted
from the ALFALFA dataset, traced by the dashed blue line.
In March 2012, a single-pixel spectrum was obtained using 
the Arecibo telescope and the `L-band-wide' (LBW) receiver 
with a velocity resolution of 1.2 \kms ~(after Hanning smoothing); 
the on-source integration time was 3 minutes. 
This spectrum is displayed as the solid red line in Figure \ref{fig:f1};
it matches and corroborates the ALFALFA spectrum. Since a single pointing
recovers practically all the flux of the ALFALFA map, the source
angular diameter must be significantly smaller than the FWHM=4\arcmin~ Arecibo beam;
however, close inspection of the ALFALFA map suggests that the Arecibo beam
starts to resolve a weak extension of the source to the SE. It is unlikely
to be due to telescope beam sidelobes. Thus, a rough
estimate of the half-mass HI radius between 0\arcmin.8 and 1\arcmin.5 can be made.
No indication of a narrow component of width less than 10 \kms~ is 
evident in either line profile.

\begin{figure}
\figurenum{2}
\begin{center}
\includegraphics[width=\linewidth]{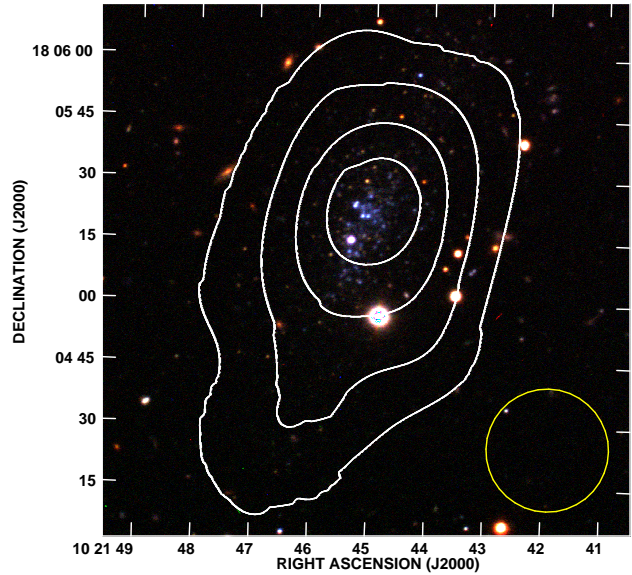}
\caption{VLA-C integrated HI synthesis preliminary image of Leo P, smoothed to an
angular resolution of 30\arcsec~ (shown as the yellow circle
in the lower right), superimposed on the WIYN optical image of 
\citet{Rhode2013}. HI column density contours are shown at 
$0.5\times 10^{20}$ cm$^{-2}$ intervals, starting at $0.5\times 10^{20}$ cm$^{-2}$
for the outermost contour.
}
\end{center}
\label{fig:f2}
\end{figure}

In April 2012, 5 hours of VLA observations in the C configuration 
were obtained as part of the Observatory 
Director's Discretionary Time. A resulting HI image, smoothed to an 
angular resolution of 30\arcsec,  is shown in the form of contours of HI 
column density overlaid on the WIYN 3.5m optical image in Figure~\ref{fig:f2}.
The peak column density is $\sim$2.3\,$\times$\,10$^{20}$ cm$^{-2}$
and the HI major axis radius is 1\arcmin ~at the column density
level of 5\,$\times$\,10$^{19}$ cm$^{-2}$. However, only half
of the ALFALFA flux is recovered in the VLA image (including flux
rescaling: see {Jorsater \& van~Moorsel 1995}\nocite{jvm95}) at
30\arcsec~ resolution, indicating that much of the HI detected by
ALFALFA is in a diffuse component more extended than that scale.

Figure \ref{fig:f1} displays the VLA spectrum integrated over the 
whole source (dotted line), which is vaguely suggestive of having two 
spectral components: one as broad as indicated by the Arecibo
observations, and a second, narrower than 10  \kms~ FWHM. If confirmed, 
such a bimodality could be attributed to the
presence of two thermally stable phases of the atomic gas: a cold (T \textless
1000 K) neutral medium (CNM) associated with the spectrally narrow
component that could arise from a single dense, star-forming region, 
and a warm (T \textgreater 6000 K), predominantly neutral medium (WNM),
enveloping the former, associated with the spectrally broad component 
(Young \& Lo 1996, Warren \etal ~2012). The relative masses of the two
components could be used to constrain thermal models of the ISM. The
extant observations cannot reliably provide such a constraint yet.
They do however provide evidence for a rotating disk. 

Smoothed to an angular resolution of 30\arcsec and a spectral resolution of 3.3 \kms, 
the VLA-C data reveal a sustained velocity gradient along the
major axis (with position angle of $34^\circ$ clockwise from North) of the 
column density distribution shown in Figure 2. While the signal-to-noise
is poor, a rotation curve can be extracted from the image. It is shown in Figure 3. 
No ordered velocity gradient is evident along the minor axis. The rotational 
velocity amplitude $V_{rot}$ that can be measured
from Figure 3, uncorrected for orbital plane inclination, is $9.0\pm 1.5$ \kms. Assuming that
disk inclination can be inferred from the axial ratio of the outermost contour in Figure 2, 
after correction of that ratio for beam smearing, we obtain $\sin i \simeq 0.85$
(the axial ratios of the three lowest column density contours in Figure 2 are
consistent with each other within an estimated accuracy of 20\%). The
$V_{rot}$ corrected for inclination is then 10.6 \kms. Thermal broadening and 
large-scale motions appear to contribute comparably to the linewidth of the integrated 
spectrum.

The preliminary VLA-C data discussed here are soon to be complemented by
additional, approved observations in array configuration D 
(Cannon \etal, in preparation)
and by observations with the GMRT array (Chengalur \etal, in preparation).

Table \ref{tab:parms} summarizes the basic parameters of Leo P as follows:
\begin{itemize}
\itemsep 0pt
\item Coordinates of the HI emission centroid, for both Arecibo and the VLA maps; 
      the source of all parameters labelled 'AO' in the table refer to the ALFALFA data set
      and those labelled `VLA' refer to the VLA-C configuration data 
\item $V_\odot$ is the AO heliocentric velocity;
\item $V_{gsr}=V_{lsr}+225\sin l\cos b$ is the radial velocity in the Galactic
      Standard of Rest frame and $V_{lsr}$ is that in the Local Standard of Rest, 
      with an assumed solar motion of 20 \kms~ towards $l=57^\circ$, $b=25^\circ$;
\item $V_{LG}$ is the radial velocity in the LG dynamical rest frame, 
      with respect to which the MW motion is of 316 \kms~ towards an apex of Galactic 
      coordinates $(l,b,)=(93^\circ, -4^\circ)$ \citep{Karachentsev1996};
\item $W50$ is the AO velocity width, measured at half power; 
\item $F_{HI}$ is the integrated flux density under the line of the AO feature;
\item $R_{HI}$ is an (uncertain) estimate of the HI radius in kpc,
      corresponding to 1\arcmin~ in angular size along the major axis
      of the isophote at $5\times 10^{19}$ HI atoms cm$^{-2}$ shown 
      in Figure 2; the radius corresponding to the isophote at the level of half the 
      peak HI column density is one-half the value of $R_{HI}$ in Table 1;
\item $M_{HI}$ is the HI mass, in solar units, as derived from $F_{HI}$;
\item $V_{rot}$ is the maximum observed rotational velocity, corrected for disk inclination,
      as extracted from Figure 3;
\item $M_{dyn}(<R_{HI})$ is an estimate of the upper limit of the 
      dynamical mass within the HI radius, on the assumption that the object is a
      self--gravitating system, computed via
      \be
      M_{dyn}(<R_{HI}) \simeq R_{HI} \sigma^2/G \label{eq:mdyn}
      \ee
      with  $\sigma=W50/2\sqrt{2\ln 2}$. The same result is obtained if $\sigma$ is 
      replaced by $V_{rot}$;
\item $M_{HI}/M_*$ is the ratio of the HI mass to the stellar mass; the latter
      was estimated by \citet{Rhode2013}. The assumed error is not driven by
      uncertainties in the photometry, but rather by the assumed protocol to convert
      optical flux and color into mass.
\end{itemize}

\begin{deluxetable}{lcc} 
\tablewidth{0pt}
\tabletypesize{\scriptsize}
\tablecaption{Properties of Leo P \label{tab:parms}}
\tablehead{
\colhead{Parameter}& \colhead{Value}    
}
\startdata
Right Ascension (J2000)	AO	& 10:21:45.0 	\\
Declination (J2000)	AO	& 18:05:01   	\\
Right Ascension (J2000)	VLA	& 10:21:44.8 	\\
Declination (J2000)	VLA	& 18:05:20 	\\
Galactic longitude		& $219^\circ.654$  \\
Galactic latitude		& $54^\circ.430$   \\ 
$V_\odot$ (\kms)  AO	 	& $264\pm2$	\\
$V_{gsr}$ (\kms)		& $177$		\\
$V_{LG}$  (\kms)		& 137		\\
$W50$ (\kms)	AO		& $24\pm 2$      \\
$F_{HI}$ (Jy \kms)		& $1.31\pm 0.04$ \\
$R_{HI}$ (kpc)  VLA  		& $0.29\pm 0.07 D_{Mpc}$ 	 \\
$M_{HI}/M_\odot$ AO	        & $3.1\times 10^5 D_{Mpc}^2$	\\
$V_{rot}$ (\kms) VLA            & $10.6\pm 2.2$ \\ 
$M_{dyn}(<R_{HI})/M_\odot$      & $8\times 10^6 D_{Mpc}$ 	\\
$M_{HI}/M_*$                   &  $2.6\pm 0.5$ \\
\enddata
\end{deluxetable}

\subsection {Preliminary Estimate of Distance}\label{BTRF}

The Tully-Fisher relation \citep{TF77}
has been effective in the determination of cosmic distances. For galaxies
with $V_{rot} > 50$ \kms, that determination's accuracy is 15--20\%. In its usual
format, as the scaling law between optical or near infrared luminosity vs.
rotational velocity $V_{rot}$, it is not useful in practice with galaxies 
with small $V_{rot}$, both because the scatter grows with decreasing 
$V_{rot}$ below 50 \kms, and the power law
behavior characterizing faster rotators fades. However, in the form
of a relation between observed baryonic mass and rotational width (hereafter 
referred to as the baryonic Tully-Fisher relation, or 'BTFR'), its power law 
behavior holds over more than five orders of magnitude 
in baryonic mass, with only moderate increase in scatter to the lowest $V_{rot}$.
\citet{McGaugh} has recently calibrated the BTFR to 
$V_{rot}\simeq 20$ \kms. A preliminary determination of the distance of Leo P can
thus be obtained, with a modest extrapolation of the BTFR.

\begin{figure}
\figurenum{3}
\begin{center}
\includegraphics[width=\linewidth]{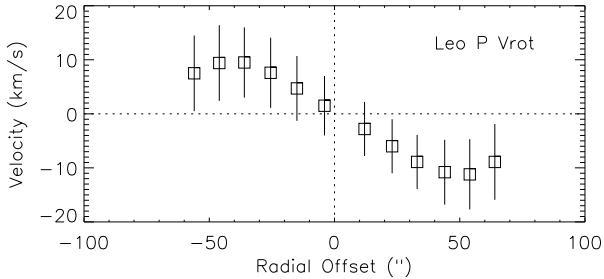}
\caption{Rotation curve of Leo P, extracted along the major axis
of the preliminary VLA-C image shown in Fig. \ref{fig:f2}, uncorrected for
disk inclination.}
\end{center}
\label{fig:f3}
\end{figure}

We define the observed baryonic mass in the same way as done by McGaugh (2012), i.e.
adding the stellar mass with the gas mass as inferred from the HI observations, multiplied
by 1.33 in order to account for Helium mass, assumed to be present in primordial
abundance. We ignore the contribution of molecular gas. The stellar mass is derived 
from the optical data by Rhode \etal (2013). Thus the observed baryonic mass is
\begin{align}
M_{bar}& = 1.33 M_{HI} + M_* = (1.33 + 2.6^{-1})~M_{HI} \nonumber \\
&=  (5.3\times 10^5~M_{\odot})~D^2_{Mpc}
\end{align}

\citet{McGaugh} shows the BTFR in his Figure 1, which is partially reproduced
in our Figure 4; the best fit power law to the data estimated by \citet{McGaugh} is
\be
\log (M_{bar}/M_{\odot}) = 2.01 + 3.82\times \log V_{rot}  \label{eq:btfr}
\ee
with a mean scatter in $\log M_{bar}$ of 0.24 dex. The scatter is however clearly 
increasing with decreasing $M_{bar}$. For a value more appropriate to the low mass 
end of the BTFR we assume 0.36 dex, 50\% higher than the mean value. The intersection 
of Leo P's $\log V_{rot}$ and the BTFR template relation yields a determination of the 
galaxy's distance: 1.3 Mpc. We coarsely estimate that the combination of errors in 
the estimate of the galaxy's disk inclination and rotational velocity is 20\%. 
Combining this with an assumed scatter about the BTFR of 0.36 dex results in an 
estimated error for the distance of 72\%, i.e. $D_{Mpc}=1.3^{+0.9}_{-0.5}$. This is
statistically in agreement with the optical estimates of \citet{Rhode2013} and 
McQuinn (private communication), discussed in section 3.1.  
We thus adopt a distance of $D_{Mpc}=1.75$ for the remainder of this paper, for
consistency in derived parameters with the papers by \citet{Rhode2013} and
\citet{Skillman}.

\begin{figure}
\figurenum{4}
\begin{center}
\includegraphics[width=\linewidth]{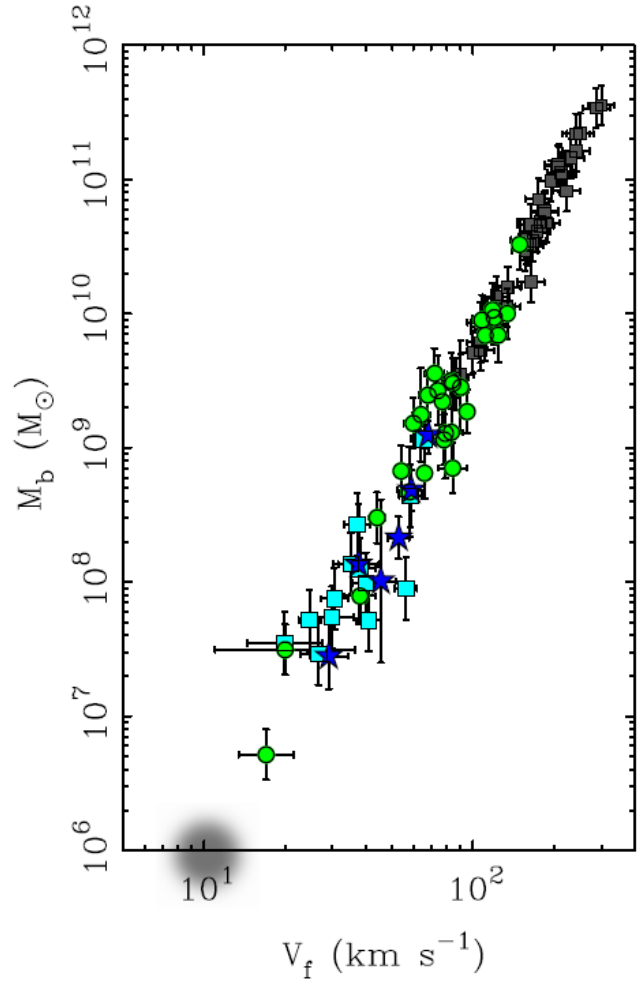}
\caption{Location of Leo P (fuzzy symbol) on the BTFR, plotted on a reproduction 
of Fig. 1 of \citet{McGaugh}, showing a template of the baryonic BTFR
of gas-rich galaxies (used with permission of the author and 
of the American Astronomical Society). The rotational velocity $V_f$ is 
measured in the same manner as our $V_{rot}$, from HI synthesis data.}
\end{center}
\label{fig:f4}
\end{figure}

\section {Discussion}\label{barcomp}

In a very useful recent compilation of galaxy properties in the LG and 
nearby groups, \citet{McConnachie2012} separates LG members into 3 subgroups:
MW satellites, M31 satellites and satellites of the LG as a whole. Galaxies
that fall within the caustic curves of constant escape velocity in the R-V
plane of each subgroup --- where R is the distance of an object from the 
barycenter of the subgroup and V that object's velocity in the subgroup's
rest frame --- are assigned membership to the subgroup. At a distance of 
1.75 Mpc, the location of Leo P in the R-V planes of each subgroup is incompatible 
with membership in any of them. At that distance, its HI mass, as derived 
from the ALFALFA data, is $1.0\times 10^6$ \msun, its stelllar mass 
$3.6\times 10^5$ \msun, its HI radius 0.5 kpc and its  dynamical mass within 
that radius $1.4\times 10^7$ 
\msun. 

It is interesting to note that the galaxy Leo I, which in terms of its sky 
location ($6^\circ .6$ angular separation) and heliocentric radial velocity 
(282 \kms vs. 264 for Leo P) would appear to be a close companion 
of Leo P, has a measured distance of 0.25 Mpc; that places Leo I within the 
300 kpc virial radius of the MW, assumed to have a mass of $10^{12}$ \msun.
Correcting its velocity for MW rotation, Leo I is traveling through the MW
corona at $\geq 174$ \kms. The ALFALFA survey yields an upper limit to 
its HI mass of $6.0\times 10^3$ \msun. The nearest neighbors of Leo P appear 
to be the galaxies in the sparse NGC 3109 group, including Sextans B, located 
at 1.4 Mpc from us and $\sim 350$ kpc from Leo P. The largest galaxy in 
that group, NGC 3109, has a V-band absolute magnitude of -14.9 and is located 
44$^\circ$ away from Leo P. The nearest giant galaxy to Leo P is the MW itself.
Leo P is a bona fide gas bearing,
star forming galaxy inhabiting a very low-mass halo, located in a low density 
environment lacking any massive companions, in the immediate vicinity of
the LG. Its measured oxygen abundance is lower than that of any galaxy in the 
Local Volume and is similar to those of I Zw 18 and DDO 68. Leo P is however 
closer to us than the latter two by 11 and 7 times respectively. We use a 
distance to DDO 68 of 12.1$\pm 0.7$ Mpc based on recent TRGB data (A. Aloisi,
personal communication). Unlike I Zw 18 and DDO 68, Leo P
shows no evidence for tidal disturbances, although the asymmetry shown by 
the SE extension of the HI isophotes in Figure 2 could be the remnant of a past
mild encounter. Forthcoming HI synthesis observations should help clarify this matter. 
There is however no evidence, either in optical or HI surveys, of any massive 
system in the vicinity of Leo P. In this object, the evolutionary history 
and extreme metallicity of a very low mass system can thus be studied in a
laboratory environment ``as clean as it gets''.

The ``dwarf galaxy problem'', the observed underabundance of dwarf
galaxies with respect to theoretical expectations, appears to be
tightly related to their baryon deficiency \citep{Papastergis2012}.
For low mass halos,
the baryon deficiency is attributable to the shallow potential well of those
systems, which makes them unable to retain their gas, which is lost due
to either heating by the metagalactic UV field or to galactic winds,
after episodic star formation events. This is shown effectively by the
simulations of \citet{Hoeft2006}, \citet{Hoeft2010} and others.
Heating of the IGM raises the Jeans mass, making IGM gas infall onto low
mass systems ineffectual, yet these circumstances may change at later 
epochs, then gas accretion and star formation in a dwarf system may resume 
\citep{Ricotti2009}. The transition from halos capable of retaining most 
of their baryons to those losing most of them takes place over a narrow 
range of halo masses. The mid--point for that transition is referred to as
the `characteristic halo mass' $M_c$. This would be the halo mass 
marking the onset of the dwarf galaxy problem. One should thus expect 
that the baryon--to--dark matter fraction of halos with mass $M_h<M_c$  
would be measurable --- if at all --- at levels much lower than the cosmic 
fraction of $\sim 1/6$. Within this scenario, the thermal models of SMW02 
are most useful in estimating the detectability of these objects. 
They indicate that halos with $M_h <10^9$ \msun ~can retain a small 
fraction of their baryons, albeit far below the cosmic baryon fraction
as N-body simulations suggest. 
Most of those baryons would be present in the form of a warm, ionized, 
thermal gas phase (WIM) with a temperature near 7000K or higher, depending
on the gas metallicity. An even smaller fraction 
of the baryon mass could be present within the inner region of  the WIM, 
in the form of a warm neutral gas phase (WNM) or even in the form of a 
cold neutral phase (CNM), eventually capable of converting some of its
gas into stars.  According to those models, the neutral gas content of 
a $10^9$ \msun ~DM halo would 
thus not be  around $10^8$ \msun, but rather a few orders of magnitude 
lower, even below $10^6$ \msun. Detection of such weak HI emitters 
would have been impossible by past or current wide field surveys 
beyond a few Mpc from us. Given their shallow potential wells, their
gas content would be of marginal stability and their detection may
require that they be located in environments sheltered from interaction with threatening neighbors and able to provide a measure of confining IGM pressure.

G+10 reported the discovery of a category of ultra-compact HI HVCs 
which are plausible minihalo candidates, as shown by the SMW02 models. 
Leo P belongs to that category of objects.
Leo P, Leo T --- the LG transition galaxy which is located 417 kpc from the MW,
just outside the MW virial radius \citep{Ryan-Weber2008} --- and the dwarf 
companion of NGC 2903 \citep{Irwin2009} may be the most extreme confirmed
examples of gas-bearing minihalos. Others may exist with even fainter
stellar counterparts, and some with no detectable starlight (or HI) at all. 
The catalog of UCHVCs as minihalo candidates of \citet{Adams2013}, extracted
from the ALFALFA data base, provides us with the targets that may help to 
shed further light on this category of dim galaxies.

\vskip 0.1in

The authors wish to thank the Director of the VLA for his positive
response to a request for discretionary time (program code 12A-456; PI
J. Cannon). We thank R. Koopmann,
P. Troischt and their student members of the Undergraduate ALFALFA
team for conducting the confirming L-band wide
observations. The Undergraduate ALFALFA team is supported by NSF grants 
AST-0724918, AST-0725267, AST-0725380, AST-0902211, and AST0903394. 
The ALFALFA work at Cornell is supported by NSF grants AST-0607007 and 
AST-1107390 to RG and MPH and by grants from the Brinson Foundation. 
EAKA is supported by an NSF predoctoral fellowship. KLR is supported by 
NSF Faculty Early Career Development (CAREER) award AST-0847109 and JMC
by NSF grant AST-1211683.

\end{document}